\documentclass[a4paper,fleqn]{cas-dc}
\usepackage[authoryear]{natbib}

\def\tsc#1{\csdef{#1}{\textsc{\lowercase{#1}}\xspace}}
\tsc{WGM}
\tsc{QE}
\usepackage{aas_macros}

\begin{document}
\let\WriteBookmarks\relax
\def\floatpagepagefraction{1}
\def\textpagefraction{.001}

\shorttitle{SIDM Core Collapse in \textsc{Hi}-Rich Galaxies}    

\shortauthors{D. Kong, H.-B. Yu}  

\title [mode = title]{Probing Signals of Self-Interacting Dark Matter Core Collapse in \textsc{Hi}-Rich Galaxies}

\author{Demao Kong}[
        orcid=0000-0003-1723-8691,
]
\credit{Conceptualization, Investigation, Formal analysis, Validation, Visualization, Writing – original draft, Writing – review \& editing}

\cormark[1]


\ead{dkong012@ucr.edu}

\author{Hai-Bo Yu}[
    orcid=0000-0002-8421-8597,
]
\credit{Conceptualization, Investigation, Resources, Supervision, Validation, Writing – review \& editing}

\ead{haiboyu@ucr.edu}

\affiliation{organization={Department of Physics and Astronomy},
            addressline={University of California, Riverside}, 
            postcode={92521}, 
            state={California},
            country={USA}}

\cortext[1]{Corresponding author}

\begin{abstract}
We analyze rotation curves of five \textsc{Hi}-rich galaxies recently discovered with MeerKAT. These galaxies exhibit sharply rising rotation curves, while their baryonic components are not dynamically dominant, suggesting that their dark matter halos have high inner densities. When fitting the standard Navarro–Frenk–White (NFW) halo model, four galaxies require extremely high halo concentrations, exceeding the cosmological median by $5\sigma$. In contrast, self-interacting dark matter (SIDM) halos in the core-collapse phase naturally account for the high densities in these galaxies. For halos with masses around $10^{11}~{\rm M_\odot}$, those in cosmic filaments exhibit concentrations consistent with the cosmological average, while halos in cosmic nodes show relatively higher concentrations that align with the SIDM fits but remain insufficient for the NFW fits. Our analysis indicates that these \textsc{Hi}-rich galaxies may have formed in cosmic nodes of dark matter with significant self-interactions.

\end{abstract}


\maketitle
\section{Introduction}
\label{sec:intro}

The rotation curves of spiral galaxies provide essential tests for our understanding of galaxy formation, evolution, and the nature of dark matter. In the standard cold dark matter (CDM) model, the halos follow a universal Navarro-Frenk-White (NFW) density profile~\citep{1997ApJ...490..493N}, which has a cusp towards the central regions $\rho\propto r^{-1}$. However, measurements of the galactic rotation curves suggest that distributions of dark matter in spiral galaxies are more diverse than expected from the NFW profile~\citep[e.g.,][]{2015MNRAS.452.3650O,2019PhRvX...9c1020R, 2020A&A...643A.161D}. Many dark matter-dominated dwarf galaxies favor a cored halo with $\rho\propto r^0$~\citep[e.g.,][]{Flores:1994gz,Moore:1994yx,deBlok:2001hbg,deBlok:2001rgg,Salucci:2007tm,deBlok:2008wp,Oh:2015xoa}. Within the CDM model, baryonic feedback processes associated with galaxy formation and evolution could lead to the formation of a shallow density core rather than a cusp~\citep[e.g.,][]{2010Natur.463..203G, 2012MNRAS.421.3464P,Read:2015sta,Chan:2015tna,Fitts:2016usl}. These feedback effects may result in dark matter distributions in galaxies that are more diverse than those expected from the NFW profile~\citep[e.g.,][]{DiCintio:2013qxa,Tollet:2015gqa, 2017MNRAS.465.4703K,2018MNRAS.473.4392S,Lazar:2020pjs}, though more work is needed to explore if the CDM model with feedback can account for the full range of the diversity of galactic rotation curves~\citep[e.g.,][]{2020MNRAS.495...58S,2020JCAP...06..027K}.

Alternatively, self-interacting dark matter (SIDM) has been proposed to address the diversity problem; see~\citet{Tulin:2017ara} for a review. Collisional thermalization among dark matter particles in the halo can modify its structure, with details depending on the stage of gravothermal evolution. In the core-expansion phase, the halo will develop a shallow density core~\citep[e.g.,][]{2000PhRvL..84.3760S,Rocha:2012jg,2012MNRAS.423.3740V,Zavala:2012us,Nadler:2020ulu,Zhang:2024ggu}. When the halo enters the core-collapse phase, its inner density will increase and can become even higher than its CDM counterpart~\citep[e.g.,][]{2002ApJ...568..475B,Koda:2011yb,2019PhRvL.123l1102E,Nishikawa:2019lsc,Sameie:2019zfo,Turner:2020vlf,Zeng:2021ldo,Correa:2022dey,2023ApJ...949...67Y,2023ApJ...958L..39N,Fischer:2023lvl,Zhang:2024fib,Nadler:2025jwh}. Additionally, the presence of baryons in galaxies will speed up the gravothermal evolution of the halo~\citep{Feng:2020kxv,2023MNRAS.526..758Z}. The core size shrinks and density increases when the baryons become dynamically important~\citep{Kaplinghat:2013xca,2018MNRAS.476L..20R,Despali:2018zpw,Sameie:2021ang,2023MNRAS.521.4630J,Kong:2024zyw,Ragagnin:2024deh,Correa:2024vgl}. Taking these effects together, SIDM generally predicts more diverse dark matter distributions in galaxies than CDM, in accord with the observations~\citep{2017PhRvL.119k1102K,2017MNRAS.468.2283C,2019PhRvX...9c1020R,Zentner:2022xux,Yang:2023stn,2024arXiv240715005R}. 

Since the gravothermal collapse of dark matter halos is a distinctive prediction of SIDM, a key test is to investigate dark matter-dominated galaxies with sharply rising rotation curves. Recently,~\citet{2024MNRAS.529.3469G} reported a serendipitous discovery of $49$ \textsc{Hi}-rich galaxies in a $2.3$-hour open time observation with MeerKAT. Among these galaxies, six have sufficient spatial resolution to obtain their rotation curves and generate mass models~\citep{2024MNRAS.529.3469G}. Five of the six galaxies have sharp rising rotation curves, while the baryons, including stars and gas, are not dynamically important. There is no evidence that the five galaxies interact with other galaxies, causing disturbance. Therefore, they may provide intriguing tests for the core-collapse scenario of SIDM.   

In this work, we consider the five galaxies ID10, ID18, ID19, ID24, and ID48~\citep{2024MNRAS.529.3469G} and analyze their rotation curves within CDM and SIDM halo models using the Markov chain Monte Carlo (MCMC) method. We will show for the NFW halo model, its required halo concentration needs to be extremely high to fit the data, $\sim5\sigma$ above the cosmological median (except ID10), while the contraction effects on the halo profile induced by the baryons are negligible. The SIDM fits require a core-collapse solution, and the favored halo concentration is $\sim3\sigma$ above the median, and it can be lower mildly after taking into the impact of the baryons on SIDM halo evolution, as we will demonstrate using controlled N-body simulations.  Additionally, we will show that the SIDM fits align with relatively higher concentrations of dark matter halos formed in cosmic nodes, which remain insufficient for the NFW fits. 

The rest of the paper is organized as follows. In Sec. \ref{sec:dmprofiles}, we discuss CDM and SIDM halo models used in our study. In Sec.~\ref{sec:rotcurve}, we present the method of modeling rotation curves and detailed fits. In Sec.~\ref{sec:para}, we show the dark matter halo parameters inferred from the fits and the gravothermal phase of the SIDM halos. In Sec.~\ref{sec:nbody}, we present controlled N-body simulations for the galaxy ID19. In Sec.~\ref{sec:diss}, we discuss environmental effects on halo concentration. Finally, we conclude in Sec.~\ref{sec:con}.

\section{Dark Matter Profiles}
\label{sec:dmprofiles}

To fit the rotation curves, we consider both CDM and SIDM halo models for comparison. For the former, we take NFW profile \citep{1997ApJ...490..493N}:
\begin{align}
\label{eqn:nfwdensity}
    \rho_{\mathrm{NFW}}(r)=\frac{\rho_s r_s^3}{r\left(r+r_s\right)^2},
\end{align}
where $\rho_{s}$ is the scale density and $r_{s}$ is the scale radius. The mass profile is given by
\begin{align}
\label{eqn:nfwmass}
   M_{\mathrm{NFW}}(r)&=4 \pi \rho_s r_s^3\left[\ln \left(1+\frac{r}{r_s}\right)-\frac{r}{r+r_s}\right]. 
\end{align}
Additionally, we will use the following relations:
\begin{equation}
\label{eqn:nfwrhosrs}
\rho_s=\frac{c_{200}^3}{3 g\left(c_{200}\right)} 200 \rho_{\mathrm{crit}}(z),~r_s=\frac{r_{200}}{c_{200}},
\end{equation}
\noindent where $\rho_{\rm crit}(z)$ is the critical density of the universe at redshift $z$, $r_{200}$ is the radius where the enclosed mean mass density reaches $200 \rho_{\rm crit}(z)$, $c_{200}$ is the halo concentration parameter, and $g\left(c_{200}\right)=\ln \left(1+c_{200}\right)-c_{200} /\left(1+c_{200}\right)$. We calculate the total mass of the halo $M_{200}$ as $M_{200}=M_{\rm NFW}(r_{200})$. Our MCMC analysis takes $M_{200}$ and $c_{200}$ as input parameters to specify an NFW halo. For each galaxy, we take its redshift information from Table 2 of~\citet{2024MNRAS.529.3469G}, ranging from $z\sim0.02\textup{--}0.05$, when applying the relations in Eq.~\ref{eqn:nfwrhosrs} for the MCMC fits.

For SIDM, we use the parametric SIDM halo model~\citep{Yang:2023jwn,Yang:2024uqb}. It assumes that the density profile of an evolving SIDM halo at a given moment can be expressed in terms of a universal function form. In this work, we take the version based on the Read profile~\citep{2016MNRAS.462.3628R}  
\begin{equation}
\rho_{\text {SIDM}}(r)=f^n\rho_{\mathrm{NFW}}+\frac{n f^{n-1}\left(1-f^2\right)}{4 \pi r^2 r_c} M_{\mathrm{NFW}},
\end{equation}
where $n$ is a dimensionless parameter, $r_{c}$ is the core radius, and $f(r)=\tanh \left(r / r_c\right)$. {\cite{2024JCAP...02..032Y} fixed $n = 1$ by calibrating the Read profile to match the density profile of a simulated SIDM halo}. The time evolution of the model parameters $r_{c}$, $\rho_{s}$, and $r_{s}$ are characterized by the following relations.

\begin{multline}
\frac{\rho_s}{\rho_{s, 0}}  =1.335+0.7746 \tau+8.042 \tau^5  -13.89 \tau^7 \\ + 10.18 \tau^9+(1-1.335)(\ln 0.001)^{-1} \ln (\tau+0.001), \\
\frac{r_s}{r_{s, 0}}  =0.8771-0.2372 \tau+0.2216 \tau^2 \\ -0.3868 \tau^3+ (1-0.8771)(\ln 0.001)^{-1} \ln (\tau+0.001),   \\
\frac{r_c}{r_{s, 0}}  =3.324 \sqrt{\tau}-4.897 \tau +3.367 \tau^2 \\ -2.512 \tau^3+0.8699 \tau^4,
\end{multline}
\noindent where the subscript “$0$” denotes the corresponding value of the initial NFW profile, $\tau$ is the normalized timescale characterizing the phase of gravothermal evolution of the halo. It is calculated as $\tau = t/t_{c}$, where $t$ is the SIDM evolution duration of a galactic halo and $t_{c}$ is the core-collapse timescale. For a halo with an initial NFW profile, $t_{c}$ is given by~\citep{2002ApJ...568..475B, 2015ApJ...804..131P, 2019PhRvL.123l1102E}

\begin{equation}
\label{eqn:tc}
t_{\mathrm{c}}=\frac{150}{C} \frac{1}{\left(\sigma_{\rm eff} / m\right) \rho_{{s,0}} r_{{s,0}}} \frac{1}{\sqrt{4 \pi G \rho_{{s,0}}}},
\end{equation}
where $C=0.75$ is a numerical factor calibrated from N-body simulations and $\sigma_{\rm eff}/m$ is the effective self-scattering cross section per unit mass for a given halo~\citep{2022JCAP...09..077Y}. { We have $\sigma_{\rm eff}/m\propto t^{-1}$ for a given normalized timescale $\tau$. For a longer SIDM evolution duration, a smaller cross section is required to achieve the same density profile, and vice versa. Since the scaling relation between $\sigma_{\rm eff}/m$ and $t$ is straightforward, we will fix $t=10~{\rm Gyr}$ for all galaxies in our SIDM fits as motivated by their redshift measurements.}

In our MCMC fits, we will fix $\sigma_{\rm eff}/m=10~\rm{~cm^{2}/g}$ (SIDM10) with the following considerations. The galaxies in the sample have a maximum circular velocity of $V_{\rm max}\sim100~\rm{km/s}$, which corresponds to a halo mass of $\sim10^{11}~{\rm M_\odot}$. For such a halo to collapse within $10~{\rm Gyr}$, a cross section of $\sigma_{\rm eff}/m\gtrsim 10~\rm{~cm^{2}/g}$ is typically needed, unless the concentration is significantly higher than the cosmological median~\citep[e.g.,][]{2019PhRvL.123l1102E,Sameie:2019zfo,2023ApJ...958L..39N}, as we will discuss later. 

For the galaxies we consider, dark matter dominates the dynamics, and a cross section larger than $10~{\rm cm^2/g}$ would also give rise to an equally good fit because of the degeneracy between $\sigma_{\rm eff}/m$ and $c_{200}$~\citep{2017PhRvL.119k1102K,2019PhRvX...9c1020R,2024arXiv240715005R}. However, for galaxies that are dominated by baryons, $\sigma_{\rm eff}/m\gg10~{\rm cm^2/g}$ could make the central density too high to be consistent with observations because the core-collapse timescale can be significantly shortened by the baryons~\citep[e.g.,][]{Correa:2024vgl}. Additionally, $\sigma_{\rm eff}/m\sim10~\rm{~cm^{2}/g}$ for $V_{\rm max}\sim100~\rm{km/s}$ aligns with the velocity-dependent SIDM models that are proposed to diverse dark matter distributions of other galactic systems~\citep[e.g.,][]{Correa:2020qam,Turner:2020vlf,2023ApJ...949...67Y,2023ApJ...958L..39N,Slone:2021nqd}.  

{ To further illustrate these points, we perform a preliminary MCMC fit to galaxy ID19, allowing the cross section to vary as a free parameter. The posterior distribution between the cross section and concentration exhibits a strong anti-correlation (see Appendix~\ref{appex:c200sigmam}). As the cross section increases (decreases), the required concentration decreases (increases) accordingly for the halo to enter the collapse phase at $t=10~{\rm Gyr}$. The inferred median values of $\sigma_{\rm eff}/m$ and $c_{200}$ are approximately $10~{\rm cm^2/g}$ and $3\sigma$ higher than the cosmological concentration median, respectively. Therefore, the assumption of $\sigma_{\rm eff}/m=10~{\rm cm^2/g}$ is well justified and does not result in loss of generality. Intriguingly, the inferred concentration is consistent with the expected values for halos in cosmic nodes, as we discuss later.} 

We will also use controlled N-body simulations to test the robustness of our fits. Although the baryons are not dynamically significant for the galaxies we consider and the adiabatic contraction effect~\citep{Blumenthal:1985qy,2004ApJ...616...16G} is negligible for the CDM halo, they do have a mild impact on the SIDM halo, as we will show in Sec.~\ref{sec:nbody}.

\section{Modeling the Rotation Curves}
\label{sec:rotcurve}

We take the rotation curves for the galaxies ID10, ID18, ID19, ID24, and ID48 from~\citet{2024MNRAS.529.3469G} and perform a Bayesian analysis by utilizing MCMC method with the public code~\texttt{emcee} \citep{2013PASP..125..306F}. Note that we do not include galaxy ID8 in our analysis. This galaxy exhibits a low overall rotation velocity but a high baryon contribution, indicating that its dark matter fraction is, at face value, below the cosmological relative abundance of $85\%$, see Table 4 in \citet{2024MNRAS.529.3469G}; we also confirmed it in our preliminary analysis. 

 We decompose the total circular velocity as $V_{\rm rot}^2 = V_{\rm star}^2 + V_{\rm gas}^2 + V_{\rm dm}^2$. For the stellar ($V_{\rm star}$) and gas ($V_{\rm gas}$) contributions, we adopt their corresponding median values as reported in~\citet{2024MNRAS.529.3469G} as their uncertainties are relatively small and these galaxies are dark matter-dominated. For each galaxy, we sample the model parameters $M_{200}$ and $c_{200}$ with flat priors in the ranges $9 \leq \log_{10}(M_{200}/{\rm M_\odot}) \leq 13$ and $1 \leq c_{200} \leq 100$, respectively. Cosmological simulations of structure formation show that $c_{200}$ and $M_{200}$ are correlated as the concentration-mass relation~\citep{2014MNRAS.441.3359D}, but we do not impose it in our sampling.

\begin{figure*}[h!]
  \centering
    \includegraphics[width=1.8\columnwidth]{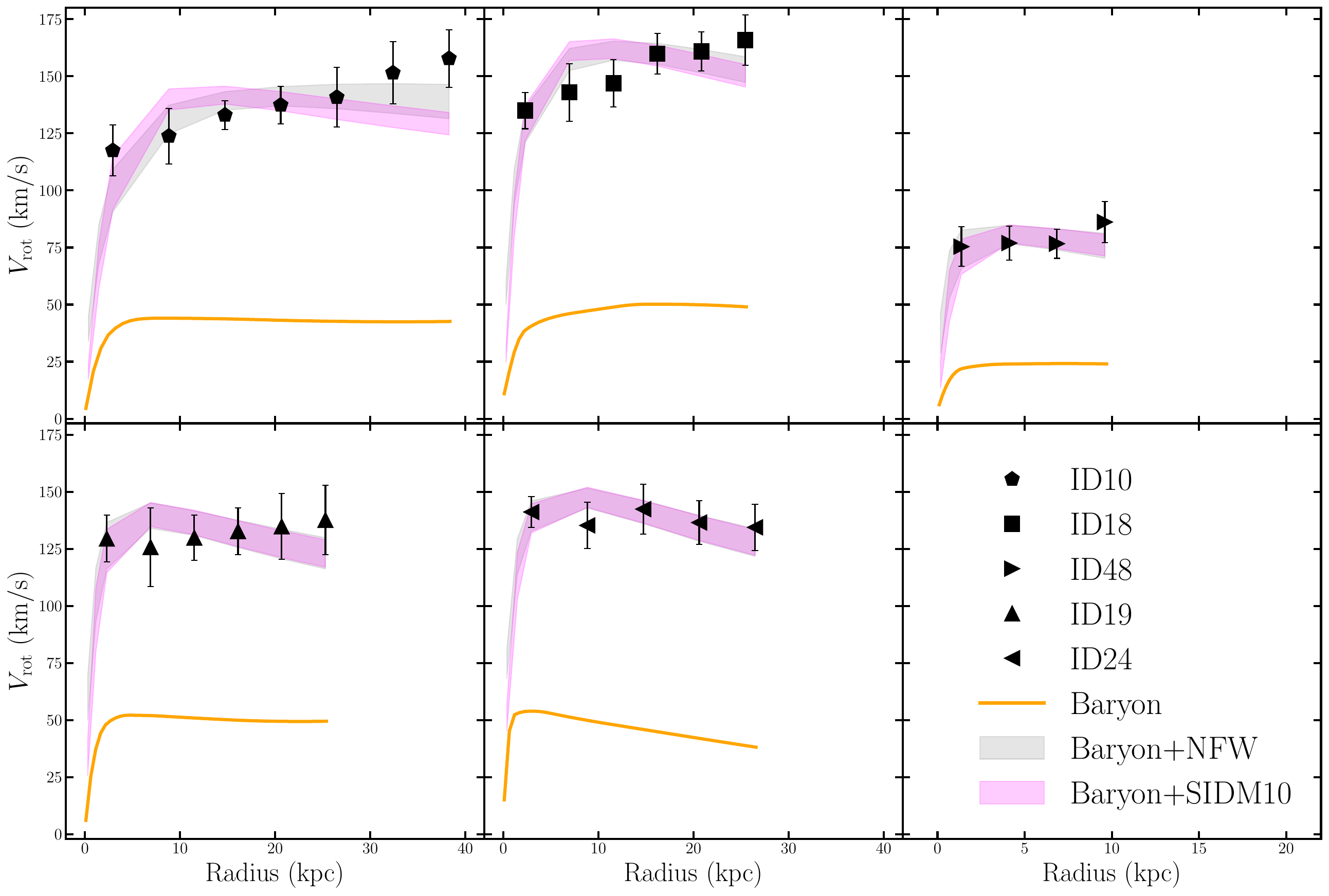}
    \caption{Rotation curves from the SIDM10 (magenta) and NFW (gray) fits, where each shaded band represents the $68\%$ confidence region spanning from the $16{\rm th}$ to the $84{\rm th}$ percentile of the MCMC chain. The black data points with error bars represent the observed total rotation curves from \citet{2024MNRAS.529.3469G}, and the orange solid curves denote the baryonic contributions (stars+gas).}
    \label{fig:fullfitband}
\end{figure*}

\begin{figure*}[h!]
  \centering
    \includegraphics[width=1.8\columnwidth]{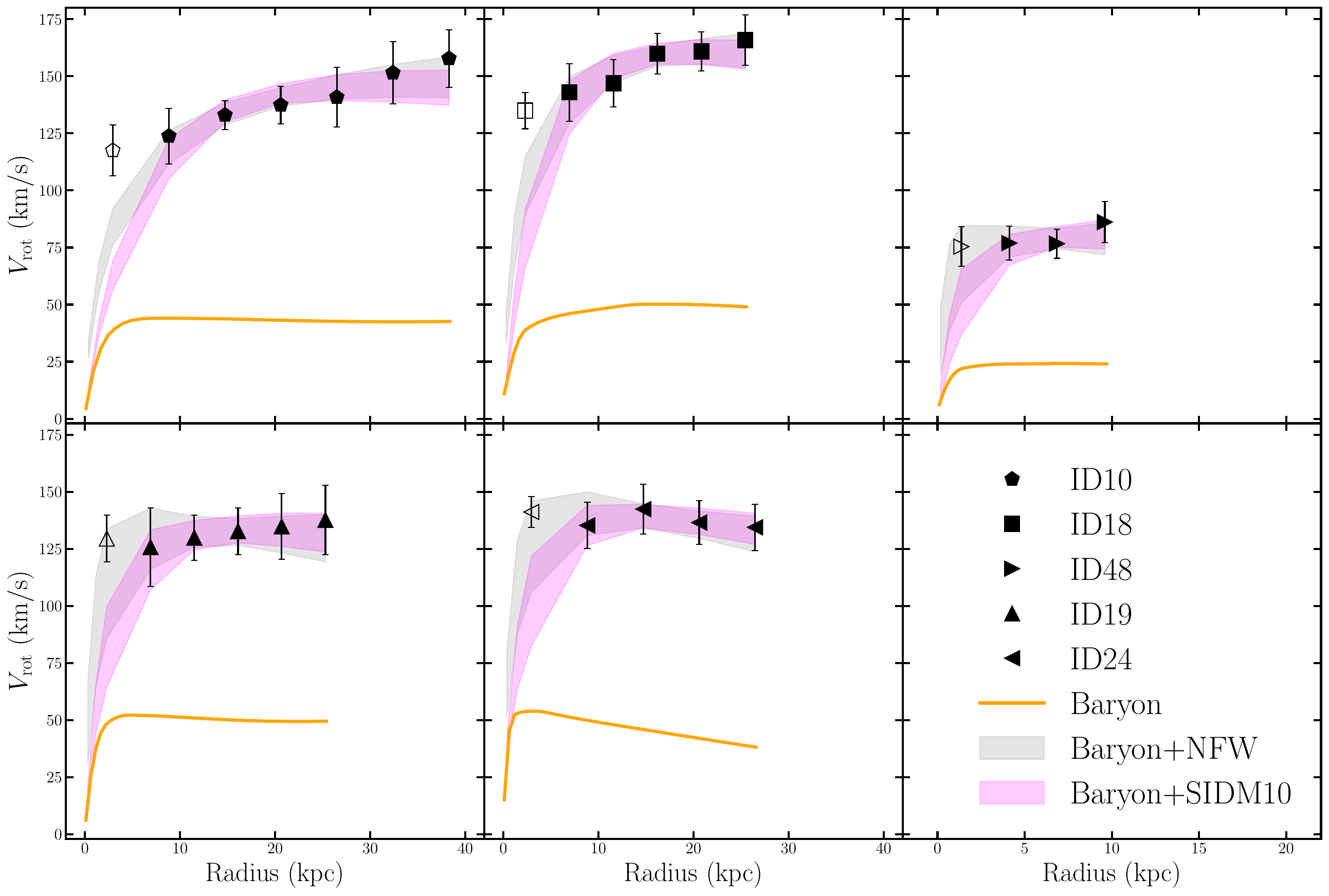}
    \caption{The same as Fig.~\ref{fig:fullfitband}, but the innermost data points (hollow markers) are excluded in the SIDM10 and NFW fits.}\label{fig:reducedfitban}
\end{figure*}

In Fig.~\ref{fig:fullfitband}, we present our MCMC fits to the rotation curves of the galaxies for the NFW (gray) and SIDM10 (magenta) halo models. The shaded regions represent the $68\%$ confidence level from the $16$th percentile to the $84$th percentile. We also show the total baryon contribution, including stars and gas (orange). Both halo models fit the data well and their predicted rotation curves overlap each other in most regions, suggesting that it is difficult to distinguish the core-collapsed SIDM and NFW halos solely from rotation curve measurements. Among the galaxies, the difference is more noticeable for ID10 as the circular velocity of the SIDM halo reaches its maximum at $r\approx10~{\rm kpc}$ and decreases slightly toward larger radii, while the NFW one is almost flattened out for $r\gtrsim10~{\rm kpc}$.

The galaxies in the sample have a sharp rising rotation curve; thus, high-density inner halos are needed. Notably, the measured circular velocities at $r\approx3~{\rm kpc}$ are already close to their peak values. To explore to what extent our fits shown in Fig.~\ref{fig:fullfitband} are driven by the innermost data points, which may be subject to relatively larger measurement uncertainties, we further fit the galaxies without including them, as shown in Fig.~\ref{fig:reducedfitban}. For ID19, ID24, and ID48, their circular velocities of the NFW halos still peaked at radii $r\approx3~{\rm kpc}$. In contrast,  the SIDM10 rotation curves rise slower for all galaxies, compared to those shown in Fig.~\ref{fig:fullfitband}, as the halos could be less collapsed without including the innermost data points, as we discuss in the next section.

\section{Halo Concentration and Gravothermal Timescale}
\label{sec:para}

\begin{figure*}[h!]
  \centering
    \includegraphics[width=2\columnwidth]{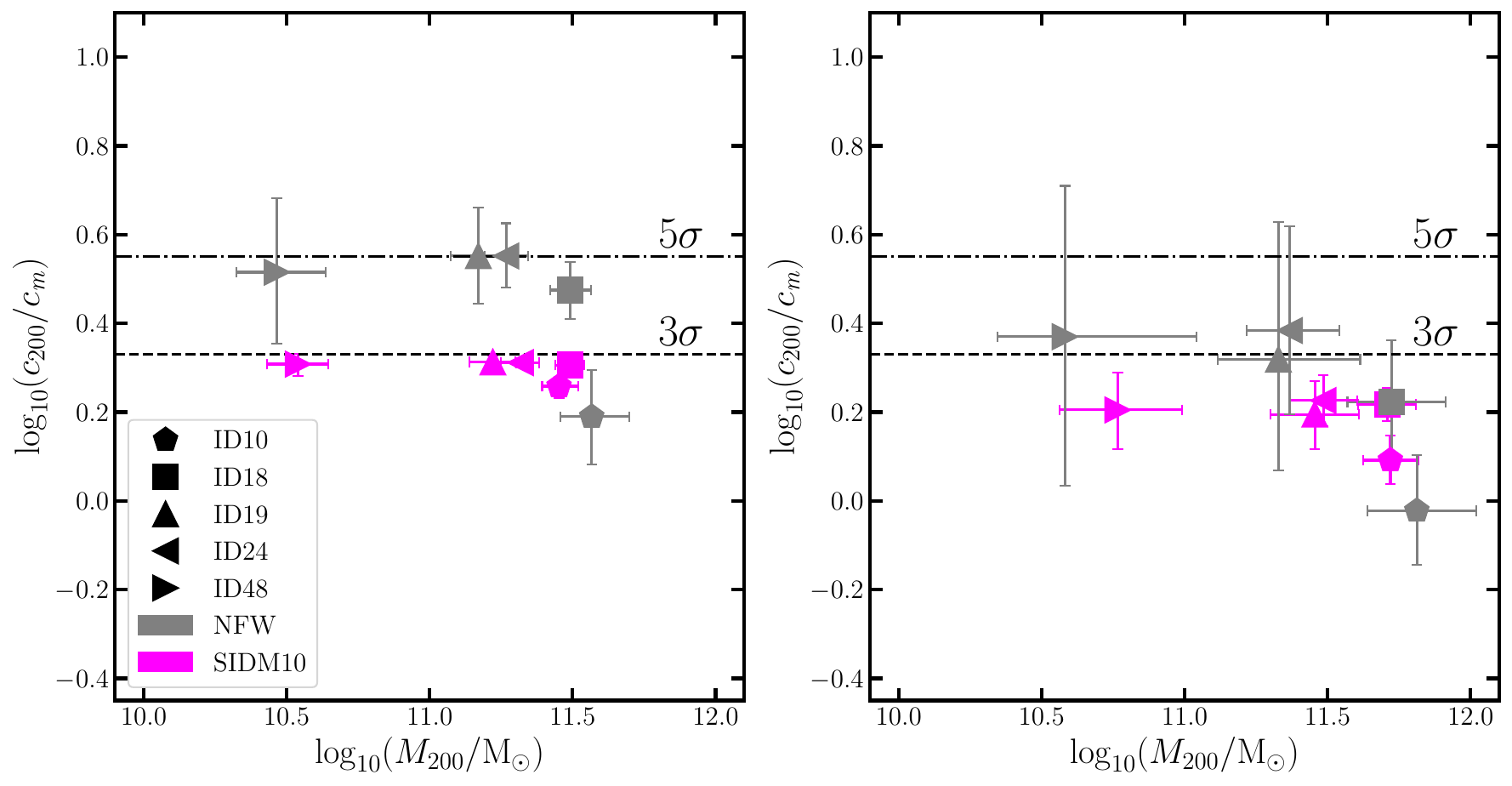}
    \caption{Inferred $\log_{10}(c_{200}/c_m)$ and $\log_{10}(M_{200})$ values for the galaxies, where the median represents the $50{\rm th}$ percentile, and the error bars indicate the $68\%$ confidence region spanning from the $16{\rm th}$ to the $84{\rm th}$ percentile of the chain. The cosmological median concentration $c_m$ is calculated using the concentration-mass relation, with $3\sigma$ (dashed) and $5\sigma$ (dash-dot) above the median shown under the assumption of a scatter of $\sigma=0.11\rm{dex}$~\citep{2014MNRAS.441.3359D}. \textbf{Left}: Fits based on the full rotation curve dataset. \textbf{Right}: Fits excluding the innermost data points.}\label{fig:paracomp}

\end{figure*}

In Fig.~\ref{fig:paracomp} (left), we show the halo concentration $c_{200}$ and mass $M_{200}$ inferred from the MCMC NFW (gray) and SIDM10 (magenta) fits with the full dataset. The median represents the $50$th percentile while the error bars represent the $68\%$ confidence level, from the $16$th percentile to the $84$th percentile. The inferred concentration value is normalized to the cosmological median $c_m$ as in~\citet{2014MNRAS.441.3359D}. For reference, the $3\sigma$ (dashed) and $5\sigma$ (dash-dot) lines above the cosmological median are shown with a scatter $0.11{\rm dex}$~\citep{2014MNRAS.441.3359D}. 

For all galaxies except ID10, the NFW fits require high $c_{200}$ values, approximately $5\sigma$ above the cosmological median. In contrast, for the SIDM fits, the $c_{200}$ values are $3\sigma$ higher from the median, while the halo masses are similar to their NFW counterparts. For the galaxy ID10, the SIDM fit prefers a slightly smaller mass but a higher concentration than its NFW counterpart. The rotation curve of ID10 is the most extended in the sample, up to $40~{\rm kpc}$. The SIDM fit tends to match the innermost data point while missing the outermost one. The trend is reversed for the NFW fit.  

Fig.~\ref{fig:paracomp} (right) shows $c_{200}$ and $M_{200}$ inferred from the fits without including the innermost data points. Compared to the results in the left panel, the NFW and SIDM fits are shifting towards lower concentration values by approximately $2\sigma$ and $1\sigma$, respectively. Meanwhile, the uncertainties increase as expected. Even when excluding the innermost points, the NFW fits still require systematically higher concentrations than the SIDM fits, except for ID10.

\begin{figure*}[h!]
  \centering
    \includegraphics[width=2\columnwidth]{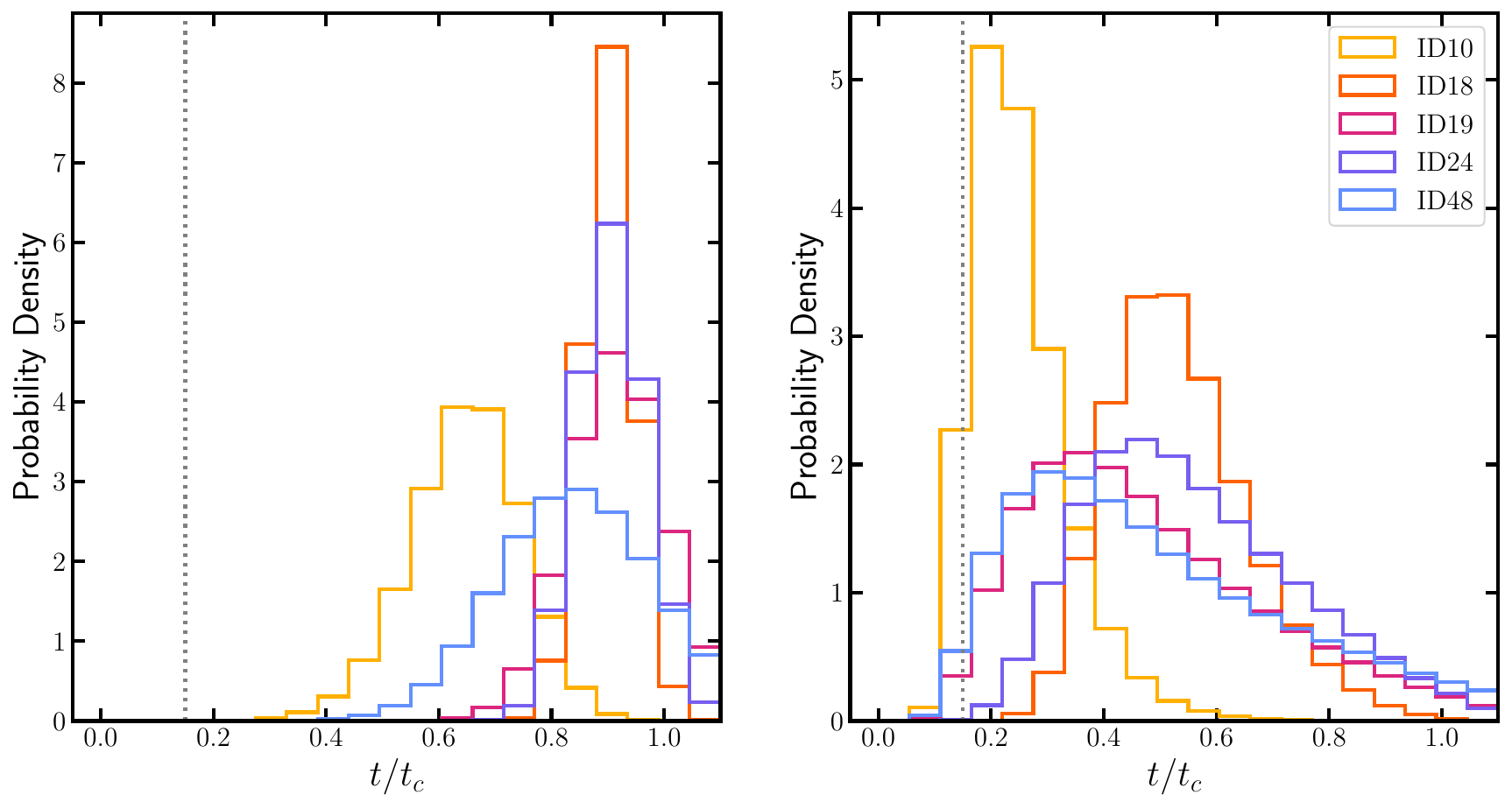}
    \caption{Histogram of the probability density for the normalized evolution timescale $\tau=t/t_{c}$ across the entire MCMC chain for the galaxies. The gray dotted line indicates $\tau = 0.146$, corresponding to the point of maximum core expansion and lowest central density for SIDM halos. \textbf{Left}: Fits based on the full rotation curve dataset. \textbf{Right}: Fits excluding the innermost data points. }
    \label{fig:tau}
\end{figure*}

Fig.~\ref{fig:tau} (left) shows the probability density for the normalized gravothermal evolution timescale $\tau = t/t_{c}$ from the SIDM fits with the full dataset. We fix $t$ to be $10~{\rm Gyr}$ and calculate $t_{c}$ using Eq.~\ref{eqn:tc} with the halo parameters from the MCMC chains. The vertical gray line denotes $\tau=0.146$ where an SIDM halo has maximum core expansion and the lowest central density accordingly~\citep{2024arXiv240715005R}. All the SIDM halos are in the collapse phase. ID10 has $\tau\sim 0.6$, while the others have $\tau \sim 0.9$, suggesting a favor towards the late phase of gravothermal evolution to fit the sharp rising rotation curves in the sample. If the fits do not include the innermost data points, the $\tau$ distribution shifts to the regime $0.2 \lesssim \tau \lesssim 0.6$ as shown in Fig.~\ref{fig:tau}~(right). The SIDM halos are at slightly earlier stages of gravothermal evolution but are still in the core-collapse phase as the maximum core expansion occurs at $\tau=0.146$. Thus, the SIDM fits favor a core-collapsed solution regardless of whether the innermost data points are included.

{ We further test the impact of measurement and modeling uncertainties on the inference of halo parameters. We perform MCMC SIDM and NFW fits with a varying stellar mass-to-light ratio and find that the inferred halo parameters are nearly identical to those from the fits with a fixed ratio, as stars are dynamically insignificant for the galaxies in our sample (see Appendix~\ref{appex:MoL}). Additionally, we fit the full rotation curves using the Einasto profile and find that the resulting concentrations are similar to those from the NFW profile fits (see Appendix~\ref{appex:ein}).}

{We also explore the effect of inclination uncertainties by varying the inclination ratio $\sin(i)/\sin(i^\prime)$ in the fits, where $\sin(i)$ is the fiducial value. To reduce the concentrations from the NFW fits ($5\sigma$) to levels comparable to those from the SIDM fits ($3\sigma$), a ratio of $0.85$ is required. A similar reduction could be achieved if the estimated distances of the galaxies were a factor of two larger than currently assumed. Future measurements will help determine whether the unusual properties of these galaxies can be attributed to measurement uncertainties.}

\section{Controlled N-body Simulations}
\label{sec:nbody}
In our MCMC fits, we did not take into account the impact of the baryons on the halo density profile. As indicated in Fig.~\ref{fig:fullfitband}, the baryonic contribution to the total measured circular velocity is less than $40\%$ for all the galaxies. For the CDM model, we use the semi-analytical formalism from~\citet{2004ApJ...616...16G} to estimate the effect of adiabatic contraction on the halo for ID19 and find that the effect leads to up to $3\%$ enhancements in the total circular velocity, compared to the NFW fit shown Fig.~\ref{fig:fullfitband}. Thus, the adiabatic contraction effect due to the baryons is negligible in the CDM model, and the halo remains a $5\sigma$ outlier in terms of its concentration. The tension would be even more exacerbated for galaxy formation models where baryonic feedback generates a density core.

In the SIDM model, the halo is more responsive to the presence of the baryonic potential because of collisional thermalization of dark matter particles~\citep{Kaplinghat:2013xca,2018MNRAS.476L..20R, 2018MNRAS.479..359S}, especially in the core-collapse regime~\citep{ 2022JCAP...05..036F, 2023MNRAS.526..758Z,Yang:2024tba}. Thus, we shall examine the impact of the baryons on the SIDM halo for the galaxies we consider. For a case study, we take the galaxy ID19, which has the highest halo concentration in the MCMC fits and the highest baryonic mass in the sample, and conduct four sets of controlled N-body SIDM simulations.

\subsection{Simulation Setup}
\label{subsec:simsetup}

We first construct a static baryonic potential at the halo center and assume that it follows the Hernquist profile~\citep{1990ApJ...356..359H} 
\begin{equation}
\rho_b(r)=\frac{\rho_{b}}{\left(\frac{r}{a}\right)\left(1+\frac{r}{a}\right)^3},
\end{equation}
where $\rho_{b}$ is the baryonic scale density and $a$ is the characteristic radius. We determine the $\rho_{b}$ and $a$ values by fitting the Hernquist profile to the circular velocity profile of stars and gas, see the Fig.~\ref{fig:fullfitband} (orange, ID19), and find $\rho_{b} = 3.82\times 10^{6}~\rm{M_{\odot}/kpc^3}$ and $a = 10.35 ~\rm{kpc}$. For the halo initial conditions, we adopt our best-fit SIDM10 model mass $M_{200} = 1.67 \times 10^{11}~\rm{M_{\odot}}$ from the fit to full dataset (Fig.~\ref{fig:fullfitband}) and $M_{200} = 2.85 \times 10^{11}~\rm{M_{\odot}}$ from the fit excluding the innermost data point (Fig.~\ref{fig:reducedfitban}). We adjust the concentration parameter $c_{200}$ by demanding the circular velocity profile of the simulated halo to match that of the MCMC best-fit model.

For the cross section, we consider $\sigma_{\rm eff}/ m = 10~\rm{cm^{2}/g}$ as in our MCMC fits. In general, there is a degeneracy among the cross section, halo concentration and mass, and baryonic potential in affecting the gravothermal evolution of an SIDM halo. By fixing $M_{200}$ and $\sigma_{\rm eff}/m$ to be their corresponding values from the MCMC fits, we investigate the degenerate effect between halo concentration and the baryonic impact. Additionally, we will conduct simulations assuming $\sigma_{\rm eff}/ m = 3~\rm{cm^{2}/g}$ with the following considerations. After implementing the baryonic potential, the velocity dispersion of dark matter particles at the halo center increases by a factor of $1.3$. Consider an SIDM model with a strong velocity dependence as $\sigma_{\rm eff}/ m \propto 1/v^{4}$~\citep{Tulin:2017ara} on the mass scale of the ID19 halo $M_{200}\approx 2\times 10^{11}~{\rm M_\odot}$ ($V_{\rm max}\approx120~{\rm km/s}$), we expect that $\sigma_{\rm eff}/m$ decreases from $10 \rm{~cm^{2}/g}$ to $\approx3 \rm{~cm^{2}/g}$.  

We use the public code \texttt{SpherIC} \citep{2013MNRAS.433.3539G} to generate the initial condition for the dark matter halos and the code \texttt{GADGET-2} \citep{2001NewA....6...79S, 2005MNRAS.364.1105S}, implemented with an SIDM module~\citep{2022JCAP...09..077Y,Yang:2020iya}, to perform N-body SIDM simulations. The total number of simulation particles is $10^{6}$, and the softening length is $1 ~\rm{kpc}$.

\subsection{Rotation Curves with the Simulated Halos}
\label{subsec:nbodyres}

\begin{figure*}[h]
  \centering
    \includegraphics[width=2\columnwidth]{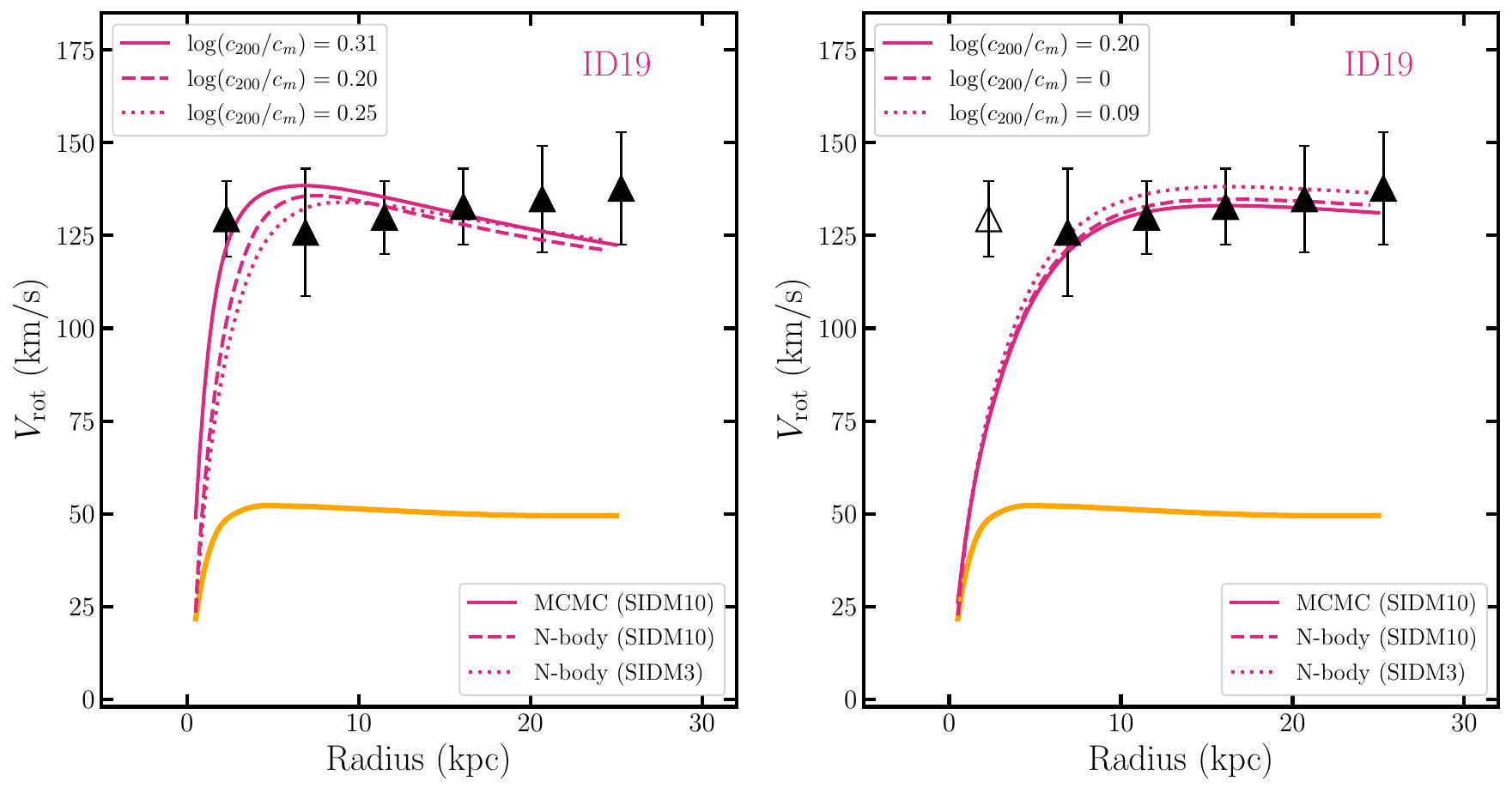}
    \caption{Rotation curves generated from the SIDM10 (dashed-magenta) and SIDM3 (dotted-magenta) N-body simulations, which account for the impact of the baryons on the halo. For comparison, the best-fit curve from the MCMC sampling (solid-magenta) and the baryonic contribution (solid-orange) are also shown. \textbf{Left}: Fits based on the full rotation curve dataset, with all three halos having $M_{200} = 1.67 \times 10^{11}~\rm{M_{\odot}}$. \textbf{Right}: Fits excluding the innermost data point, with all three halos having $M_{200} = 2.85 \times 10^{11}~\rm{M_{\odot}}$.}
    \label{fig:id19sim}
\end{figure*}

In Fig.~\ref{fig:id19sim} (left), we present the rotation curves from the SIDM10 (dashed) and SIDM3 (dotted) N-body simulations, along with the MCMC SIDM10 fit ($50$th percentile, solid), where the baryonic contribution is included (solid orange). In our simulations, we have adjusted the concentration value such that the resulting halo circular velocity agrees with the one from the MCMC fit within $10\%$. The concentration is $0.31~\rm{dex}$ above the cosmological median from the MCMC fit, while it is $0.20~\rm{dex}$ above the median from the SIDM10 simulation, which includes the baryonic impact on the halo. The baryonic potential accelerates the gravothermal evolution, and a lower concentration is needed. For the SIDM3 simulation, the required concentration is $0.25 ~\rm{dex}$ above the median between the two other cases. 

Fig.~\ref{fig:id19sim} (right) shows the rotation curves from the SIDM10 (dashed) and SIDM3 (dotted) simulations, along with the MCMC SIDM10 fit for ID19 ($50$th percentile, solid) without including the innermost data point.  Again, we see that the required concentration decreases after the impact of the baryons on the SIDM halo is included in the simulations. The SIDM10 and SIDM3 simulations suggest that a halo close to the median concentration would be sufficient if the innermost data point is excluded. 

We have seen that although the galaxies in our sample are dark matter-dominated, the baryons have a mild impact on the halo profile in the SIDM model, which further makes these galaxies less of an outlier in terms of the concentration-mass relation. In contrast, the effect of adiabatic contraction is negligible in the CDM model, and the halo must have an extremely high concentration to account for the sharply rising rotation curve. The difference is rooted in the fact that the SIDM halo is more responsive to the potential change induced by the baryons due to the collisional thermalization of dark matter particles. This feature may help distinguish the SIDM and CDM predictions with kinematic data from spiral galaxies.

\section{Cosmic Environments}
\label{sec:diss}
In CDM, the galaxies we consider are extreme outliers in terms of their halo concentration. As shown in Fig.~\ref{fig:fullfitband}, four galaxies in the sample exceed the cosmological median by $\sim5\sigma$. However, if these galaxies reside in dense environments, the standard concentration-mass relation, based on the average of simulated halos in a large cosmic volume, may not apply to them. Indeed, \citet{2024MNRAS.529.3469G} found that the $49$ \textsc{Hi}-rich galaxies can be identified with three major groups at $z\sim0.033$, $0.041$, and $0.055$, suggesting that they could have detected a super group of galaxies that formed in a dense environment. Therefore, it is interesting to explore the correlation between their sharp rising rotation curves and cosmic environments.

~\citet{Hellwing:2020fzr} showed that the concentration of filament halos increases up to $15\%$ relative to the volume average as the halo mass decreases from $7\times10^{10}~{\rm M_\odot}$ to $10^8~{\rm M_\odot}$. However, the difference is negligible for halos with masses $10^{11}~{\rm M_\odot}$, which are relevant to the galaxies studied in this work. In contrast, dark matter halos formed in cosmic nodes can have significantly higher concentration. For $M_{200}\sim10^{11}$, the concentration of the node halos vary from $c_{200}\sim15\textup{--}20$, corresponding to $2\textup{--}3\sigma$ above the median from the volume average. Intriguingly, these node halos align well with the SIDM fits but are not sufficiently dense to account for the NFW fits to the full dataset (except for ID18). Our analysis suggests that the galaxies reported in ~\citet{2024MNRAS.529.3469G} may have formed in cosmic nodes of dark matter with significant self-interactions, though further investigation is needed to validate this hypothesis.

\section{Discussion}

{ The most notable feature of this sample is that the galaxies exhibit rapidly rising rotation curves while remaining dark matter-dominated at all radii. Although the sample is small, these galaxies provide important tests for SIDM models, ruling out the core-expansion solution and favoring the core-collapse solution. Furthermore, they are also outliers in CDM. It is important to check whether the SIDM model presented in this work can also provide a good fit to other galaxies. For example,~\cite{2019PhRvX...9c1020R} performed SIDM fits on over $130$ galaxies based on the core-formation solution, assuming a cross section of $3~{\rm cm^2/g}$. A natural question is how the fits change if the cross section increases and the core-collapse solution is allowed as in this work. 

\cite{2024arXiv240715005R} partially addressed this question by fitting $14$ rotation curves of dark matter-dominated galaxies. For galaxies with slowly rising rotation curves, the fits remain nearly unchanged when the cross section increases from $3~{\rm cm^2/g}$ to $40~{\rm cm^2/g}$, with only minor changes in concentration, which is close to or below the median. This is because their halos must be in the core-formation phase, which lasts for an extended period. For galaxies with sharply rising rotation curves, their findings are well consistent with ours. Therefore, we expect that the core-formation SIDM fits in~\cite{2019PhRvX...9c1020R} are likely conservative in terms of generating diverse dark matter distributions in spiral galaxies. More work is needed in this direction, such as incorporating velocity-dependent cross sections across halos of different masses and extending the analysis to baryon-dominated galaxies. We leave this for future investigations.}

\section{Conclusions}
\label{sec:con}

We have used the NFW and SIDM halo models to fit the sharply rising rotation curves of five newly discovered galaxies, where dark matter dominates at all radii. For the NFW fits, the inferred halo concentrations are extremely high, exceeding the cosmological median by $5\sigma$ for four galaxies. In contrast, the SIDM fits, using a parametric halo model, require lower concentrations, $3\sigma$ above the median, assuming an effective cross section of $\sigma_{\rm eff}/m=10~{\rm cm^2/g}$. The SIDM halos are core-collapsed by $10~{\rm Gyr}$, resulting in high densities in the inner regions. The SIDM fits favor a core-collapsed solution regardless of whether the innermost data points are included. Additionally, the halo parameters are better constrained in the SIDM fits than in the NFW fits.

We also investigated the impact of baryons on the fits. For CDM, the effect of adiabatic contraction is completely negligible. For SIDM, controlled $N$-body simulations for one of the galaxies (ID19) show that the baryons have a mild effect in accelerating the gravothermal evolution of the halo. Including the impact of baryons slightly reduces the concentration while maintaining a similar fit. The concentrations of the node halos match those from the SIDM fits but are not sufficiently high to be compatible with those required for the NFW fits. Overall, our results indicate that the SIDM model with gravothermal collapse provides an intriguing explanation for the sharply rising rotation curves of the \textsc{Hi} galaxies. 

In the future, high-resolution measurements of the rotation curves, especially in the inner regions, can provide further robust tests on the NFW and SIDM models. It is also important to conduct kinematic measurements for more galaxies in the same cosmic region in order to have a much larger sample. On the theory side, further analyzing simulated halos formed in cosmic filaments and nodes and comparing them with the observations will be crucial to assess the challenges in the CDM model further. Additionally, it would be of great interest to conduct cosmological zoom-in SIDM simulations focusing on over-dense regions that contain cosmic nodes.

\printcredits

\bigskip
\section*{Acknowledgments}
\bigskip

This work was supported by the John Templeton Foundation under grant ID\#61884 and the U.S. Department of Energy under grant No. de-sc0008541. Computations were performed using the computer clusters and data storage resources of the HPCC at UCR, which were funded by grants from NSF (MRI-2215705, MRI-1429826) and NIH (1S10OD016290-01A1). The opinions expressed in this publication are those of the authors and do not necessarily reflect the views of the funding agencies.

\appendix

\section{Varying the Cross Section}
\label{appex:c200sigmam}

\begin{figure}[h]
  \centering
    \includegraphics[width=1\columnwidth]{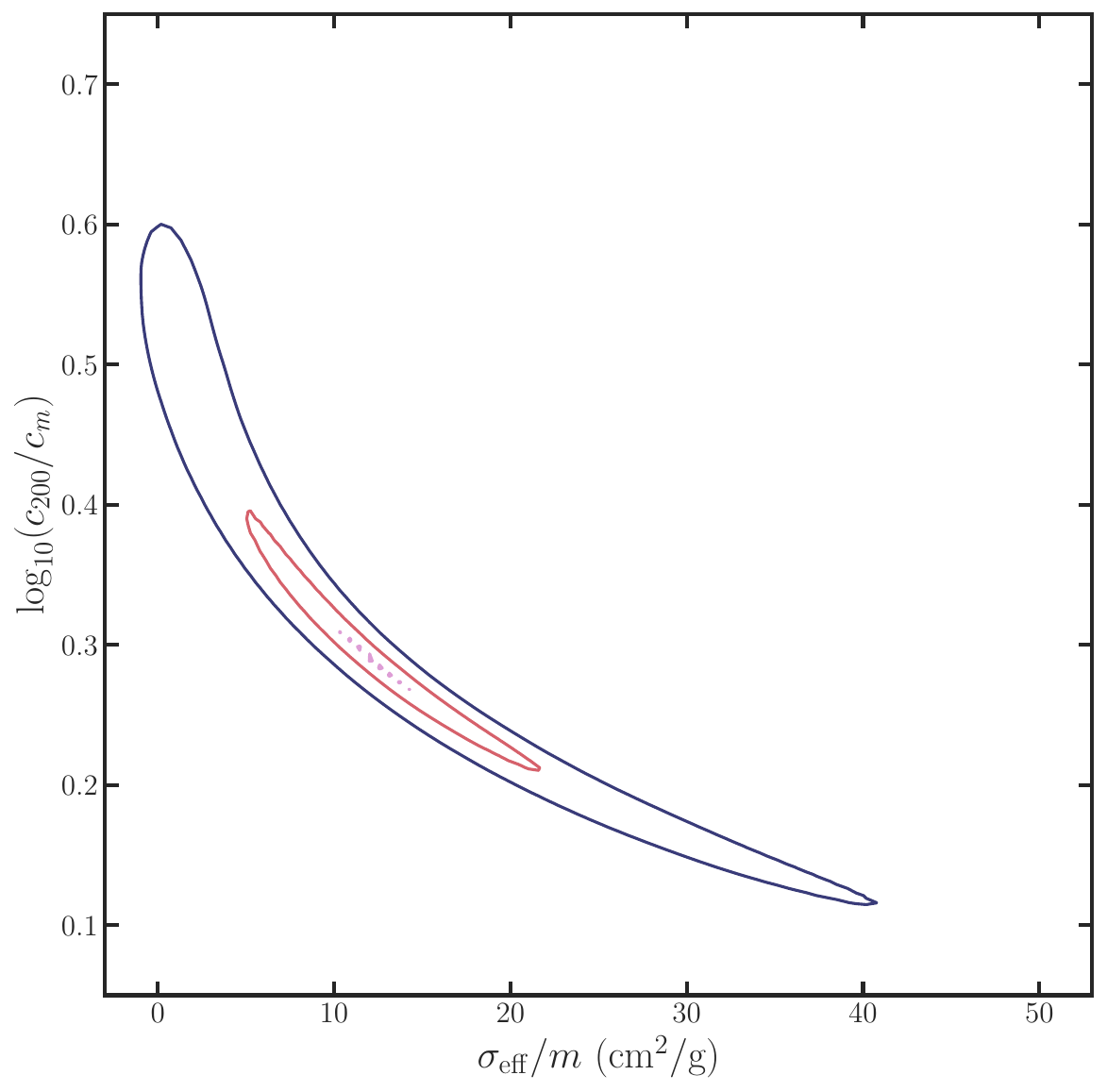}
    \caption{Posterior distributions of the halo concentration and the effective cross section for galaxy ID19. The magenta, red, and dark-blue contours represent $1\sigma$, $2\sigma$, and $3\sigma$ confidence regions, respectively.}
    \label{fig:sigma_contour}
\end{figure}

{ In our SIDM fits presented in the main text, we fixed the cross section at $\sigma_{\rm eff}/m = 10~{\rm cm^2/g}$. Here, we relax this constraint and vary the cross section with a flat prior of $0.1~\rm{cm^2/g} \leq \sigma_{\rm eff}/m \leq 100~\rm{cm^2/g}$. In Fig.~\ref{fig:sigma_contour}, we show the posterior distributions of the halo concentration and cross section for galaxy ID19, a representative case. There is a strong anti-correlation between $c_{200}$ and $\sigma_{\rm eff}/m$, which can be understood from the scaling relation $t_c\propto(\sigma_{\rm eff}/m)^{-1}c^{-7/2}_{200}M^{-1/3}_{200}$~\citep{2019PhRvL.123l1102E,2023ApJ...958L..39N}. Since the halo mass is well constrained by the outermost rotation curve data points, $\sigma_{\rm eff}/m$ and $c_{200}$ exhibit a strong degeneracy for a given collapse time $t_c$. As the cross section increases, a lower concentration is required for the halo to enter the collapse phase at $t=10~{\rm Gyr}$. The median value of $\sigma_{\rm eff}/m$ is approximately $10~{\rm cm^2/g}$, consistent with our assumption in the main results. Additionally, we note the presence of multiple local minima within the $1\sigma$ confidence regions.}

\section{Varying the Stellar Mass-to-Light ratio}
\label{appex:MoL}

\begin{figure}[h!]
  \centering
    \includegraphics[width=\columnwidth]{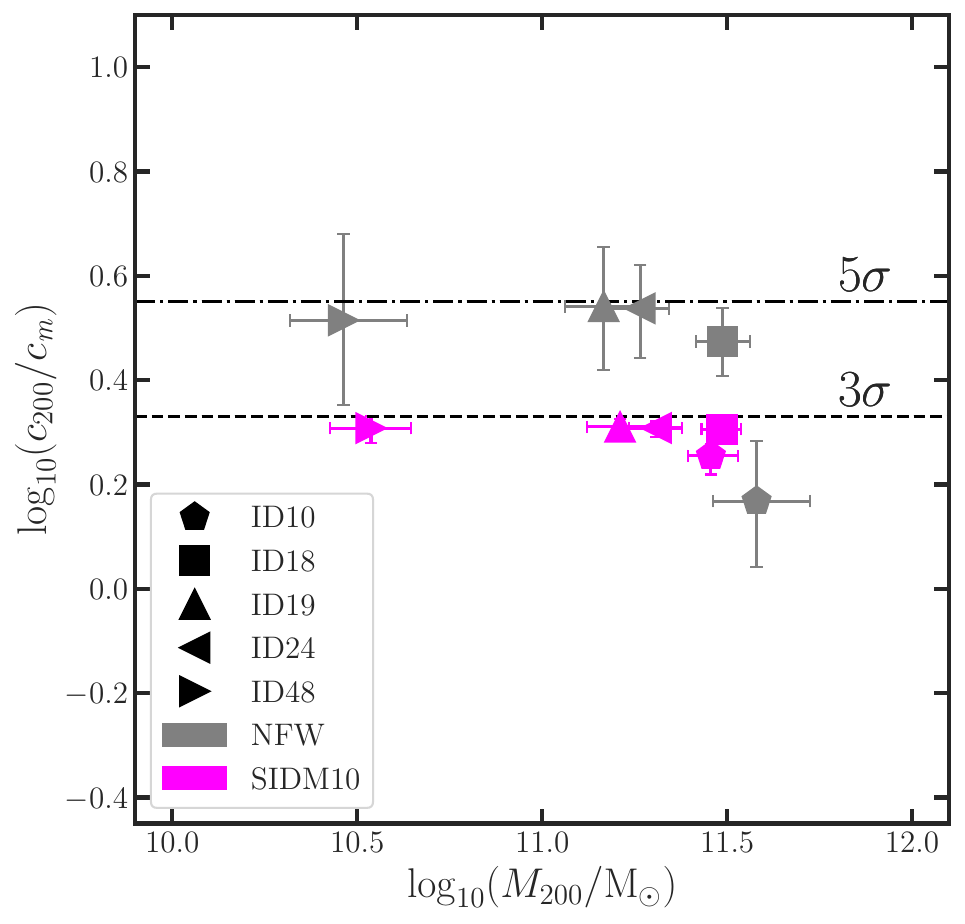}
    \caption{Same as Fig.~\ref{fig:paracomp} (left), but varying the stellar mass-to-light ratio.}\label{fig:MoLpara}
\end{figure}

{ We explore the impact of the uncertainty in the stellar mass-to-light ratio $\Upsilon^\prime_{\rm star}$ with respective to the fiducial value $\Upsilon_{\rm star}$ in~\cite{2024MNRAS.529.3469G}. We adopt a flat prior of $-0.48\rm{dex} \leq \log_{10}[\Upsilon^\prime_{\rm star}/\Upsilon_{\rm star}] \leq +0.48\rm{dex}$, which approximately corresponds to the $3\sigma$ range given the reported uncertainty $\sigma = 0.16~\rm{dex}$~\citep{2024MNRAS.529.3469G}. In Fig.~\ref{fig:MoLpara}, we present the inferred halo parameters $M_{200}$ and $c_{200}$ from the fits with a varying mass-to-light ratio. These values are nearly identical to those obtained from the fits with a fixed ratio, as shown in Fig.~\ref{fig:paracomp} (left), indicating that the stellar mass is dynamically insignificant.}

\section{Halo Parameters with the Einasto Profile}
\label{appex:ein}

\begin{figure}[h!]
  \centering
    \includegraphics[width=\columnwidth]{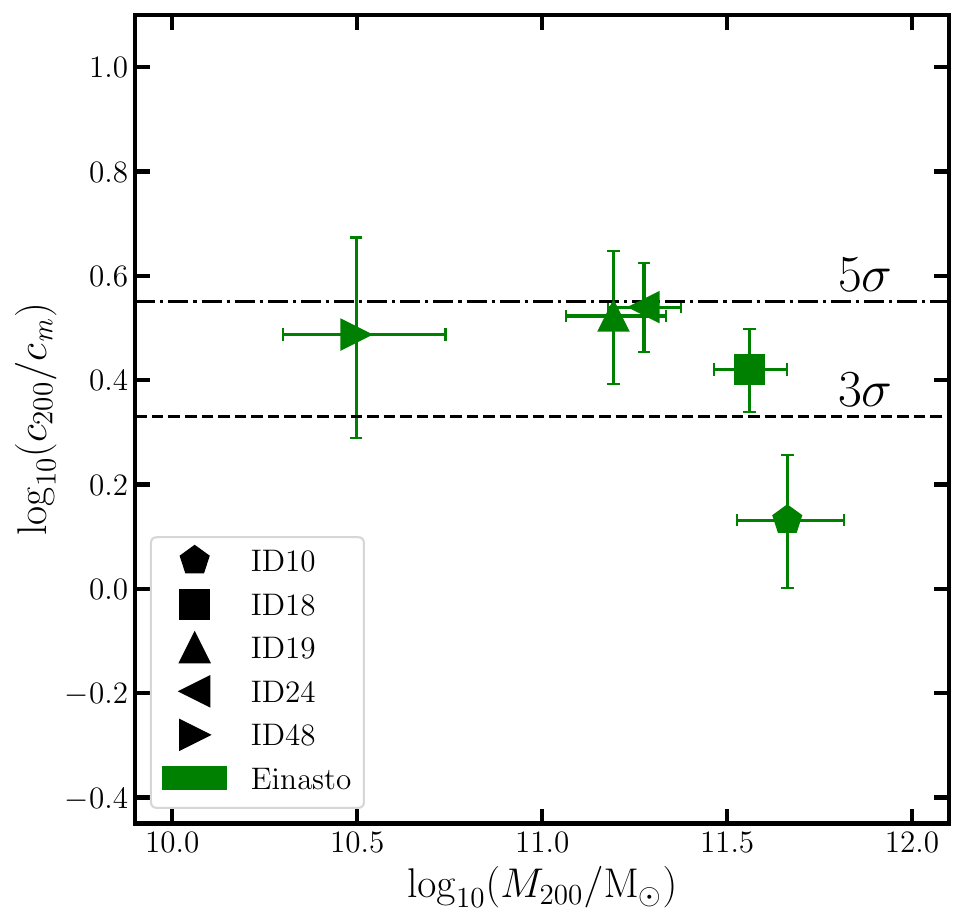}
    \caption{Same as Fig.~\ref{fig:paracomp} (left), but using the Einasto profile for fitting the full rotation curves.}\label{fig:Einpara}
\end{figure}

{ We also consider the Einasto profile~\citep{1965TrAlm...5...87E} to model the CDM halo:
\begin{equation}
\rho(r)=\rho_s \exp \left(-\frac{2}{\alpha}\left[\left(\frac{r}{r_s}\right)^\alpha-1\right]\right),
\end{equation}
where the parameter $\alpha$ controls the inner density slope of the halo. We fix $\alpha$ by adopting the relation $\alpha=0.155+0.0095 \nu^2$~\citep{2008MNRAS.387..536G}, where $\nu$ is a dimensionless parameter defined as $\nu \equiv \delta_{\mathrm{crit}}(z) / \sigma(M, z)$. $\delta_{\mathrm{crit}}(z)$ is the critical density for collapse at $z$ and $\sigma(M, z)$ is the density fluctuation within the sphere of mean enclosed mass $M$.

Fig.~\ref{fig:Einpara} shows the halo mass $M_{200}$ and concentration $c_{200}$ from the fits using the Einasto profile for the full rotation curves. The inferred halo parameters are similar to those obtained from the NFW profile fits, as shown in Fig.~\ref{fig:paracomp} (left). We have verified that while the Einasto profile is slightly denser than the NFW profile for $r\lesssim1~{\rm kpc}$, its impact on reducing the concentration value is minor.
  }

\bibliographystyle{cas-model2-names}
\bibliography{cas-refs}

\end{document}